\documentclass[aps, prb, citeautoscript, twocolumn]{revtex4-2} 

\usepackage{amsmath} 
\usepackage{amsfonts}
\usepackage{graphicx}
\usepackage{braket}
\usepackage{cancel}

\newcommand{\dd}{\mathrm{d}}
\newcommand{\ii}{\mathrm{i}}
\newcommand{\ee}{\mathrm{e}}

\setcitestyle{super,open={},close={}}

\begin{document}

\title{Surface Excitations, Energy Loss, and Decoherence in Electron Interferometry}

\author{David Kordahl}
%\author{} 
\email{dkordahl@centenary.edu} 
\affiliation{Department of Physics and Engineering, Centenary College of Louisiana, Shreveport, LA 71104}

\date{\today}

\begin{abstract}
A recent pedagogical paper by Strauch concretely demonstrated how interaction-mediated entanglement can suppress fringe visibility in a one-dimensional model of the electron double-slit experiment. Here we extend that framework to model actual experimental data from electron biprism interferometry. Kerker \textit{et al.} showed that the macroscopic QED model of Scheel and Buhmann successfully describes their measured results. We show that this Scheel-Buhmann model can be recovered from Strauch's simplified framework by employing a Markov approximation and including thermal effects. The resulting decoherence rate is expressed in terms of mode-resolved scattering probabilities familiar from electron energy-loss spectroscopy (EELS), directly relating EELS to decoherence. The thermal dependence is significant in its own right, as recent theoretical work suggests that visibility reduction could serve as a non-invasive thermal probe for nanoscale systems. This progression from a toy model, to a quantitative account of real data, to a measurement application offers a case study in how simplified models can be made experimentally relevant.
\end{abstract}

\maketitle % title page is now complete

\section{Introduction}

In electron interferometry experiments, decoherence arises from extended electromagnetic interactions with material environments, entangling a system's path degree of freedom with material excitations. This kind of interaction-based entanglement has been explored in pedagogical models emphasizing the role of subsystems \cite{Schroeder2017, Kordahl2023}, and in treatments of decoherence in interferometric geometries \cite{Kincaid-etal2016, Lerner2017}. Strauch \cite{Strauch2025} gives a particularly clear pedagogical illustration of entanglement-induced suppression of interference visibility in a simplified double-slit model. Here we show how this general approach can be extended into a quantitative account of real electron interferometry data, an example of how a simplified pedagogical model can be elaborated to become experimentally relevant.

This work is motivated by electron biprism interferometry experiments in which decoherence arises from electromagnetic coupling between an electron and a nearby dielectric surface. Sonnentag and Hasselbach \cite{SonnentagHasselbach2007} demonstrated that fringe visibility decreases as electron trajectories approach dielectric surfaces, and Kerker \textit{et al.} \cite{Kerker-etal2020} showed that fringe visibility also decreases with increasing electron path separation. Kerker \textit{et al.} tested several theoretical descriptions and found the best agreement with the macroscopic QED treatment of Scheel and Buhmann \cite{ScheelBuhmann2012}. We show that this model can be recovered from Strauch's general framework once two extensions are imposed: a Markov approximation, and the inclusion of thermal occupation effects. This derivation highlights the connection of such results to well-established models in electron energy-loss spectroscopy (EELS).

By isolating the influence of different physical effects in such calculations, we can determine which mechanisms have the most observable importance. The central result of this analysis is that the decoherence rate can be written as a weighted integral over the differential EELS probability, with the weighting factor encoding the path distinguishability of the two electron trajectories. The structure of the final expression also suggests further experimental tests, including temperature-dependent visibility measurements. Recent theoretical work by Velasco \textit{et al.} \cite{Velasco2024, Velasco2026} makes a compelling case that the predicted decline in visibility with temperature could be exploited as a novel thermal probe at the nanoscale, a suggestion with which the present analysis concurs.

The remainder of the paper is organized as follows. Sec.~\ref{sec:heuristic} introduces a schematic framework for relating fringe visibility to entanglement-induced decoherence before discussing the effects of many surface modes, the Markov approximation, and thermal effects. Sec.~\ref{sec:calculation} contrasts models for the visibility decay and connects them to EELS models. Finally, Sec.~\ref{sec:experiment} makes comparisons with published experimental results, and discusses what further tests could be carried out. Some calculational details have been relegated to an appendix. Taken together, these results illustrate how a pedagogical model can be refined into a quantitative tool, recovering an empirically successful theory and encouraging future experiments. 

\section{Visibility Calculation Structure}\label{sec:heuristic}

\begin{figure}
\centering
\includegraphics[width=0.45\textwidth]{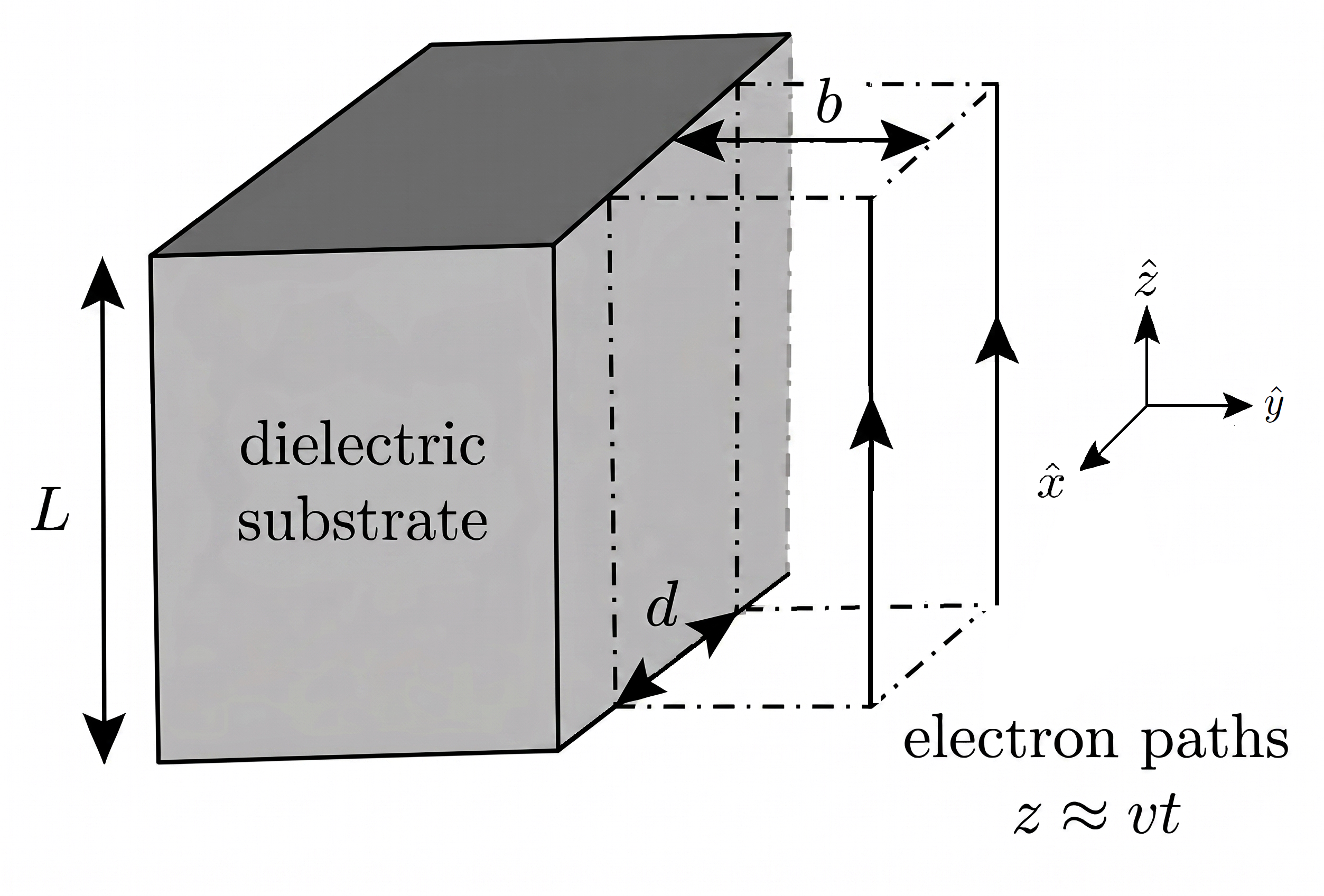}
\caption{Schematic of the interaction geometry. Two electron paths are separated by a distance $d$, propagate at a height $b$ above the substrate with velocity $v$, over a length $L$.}\label{fig:substrate-geometry}
\end{figure}

This section derives the electron fringe visibility for the geometry shown in Fig.~\ref{fig:substrate-geometry}. We adopt a Cartesian coordinate system in which a dielectric fills the half-space $y<0$, with vacuum in $y\ge0$. Two split electron trajectories propagate predominantly along the $z$ direction with velocity $v$, separated laterally by a distance $d$ along $x$ (i.e., at $x = \pm d/2$) and located at a height $y=b$ above the surface, over an interaction length $L$. 

We start with a schematic calculation modeled on Strauch's discussion of the double-slit system (Sec.~\ref{subsec:schematic}) and show how to recover the fringe visibility for the toy model with a single excited mode (Sec.~\ref{subsec:fringe_visibility}). The discussion then broadens to show how the calculation changes when many surface modes are included (Sec.~\ref{subsec:multiple_modes}), including the role of symmetry in simplifying the visibility (Sec.~\ref{subsec:symmetry}). Finally, we note the role of a Markov approximation (Sec.~\ref{subsec:Markov}) and thermal state occupation (Sec.~\ref{subsec:thermal effects}) in connecting theory with experiment.

\subsection{Schematic Calculation}\label{subsec:schematic}

Before taking on the full case, we first consider a simplified calculation. Suppose an electron has been split into two distinct paths: a \textit{right} side and a \textit{left} side, such that its $\mathbf{x}_\perp = (x,\,z)$ profile at a particular $z$ takes on the form
\begin{equation}
\ket{\Psi_e} = \frac{1}{\sqrt{2}}\left[ \psi_\text{R}(\mathbf{x}_\perp) + \psi_\text{L} (\mathbf{x}_\perp) \right].
\end{equation}
If the substrate begins in its ground state, we can write this ``initial" state of the combined system as
\begin{equation}\label{eq:initial wavefunction}
\ket{\Psi_i} = \ket{\Psi_e} \otimes \ket{0} = \frac{1}{\sqrt{2}}\left[ \psi_\text{R}(\mathbf{x}_\perp) + \psi_\text{L}(\mathbf{x}_\perp) \right] \ket{0}.
\end{equation}

As will be discussed in detail below, the substrate will have many modes, indexed on their surface wavevector, but for simplicity consider a substrate with only two possible modes, $\ket{0}$ and $\ket{1}$, representing internal degrees of freedom of the substrate. The ``final'' state of the combined system after all interactions between the electron and the substrate have effectively stopped can generically be written as
\begin{equation}\label{eq:simplified wavefunction}
\begin{split}
\ket{\Psi_f} \approx &+\frac{1}{\sqrt{2}} \psi_\text{R}(\mathbf{x}_\perp) \left[c_{\text{R}0} \ket{0} + c_{\text{R}1} \ket{1} \right] \\
&+ \frac{1}{\sqrt{2}} \psi_\text{L}(\mathbf{x}_\perp) \left[ c_{\text{L}0} \ket{0} + c_{L1} \ket{1} \right],
\end{split}
\end{equation}
where the coefficients follow the normalization conditions
\begin{equation}\label{eq:simplified normalization}
\begin{split}
|c_{\text{R}0}|^2 + |c_{\text{R}1}|^2 &= 1 \quad \text{and} \\
|c_{\text{L}0}|^2 + |c_{\text{L}1}|^2 &= 1.
\end{split}
\end{equation}

Now suppose that, after the interaction time, the wave packets $\psi_\text{R}(\mathbf{x}_\perp)$ and $\psi_\text{L}(\mathbf{x}_\perp)$ are brought to overlap (typically by the action of a downstream biprism). In this case, $\psi_\text{R}(\mathbf{x}_\perp)$ and $\psi_\text{L}(\mathbf{x}_\perp)$ are related to one another as 
\begin{equation}\label{eq:wavefunction overlap discussion}
\begin{split}
\psi_\text{R}(\mathbf{x}_\perp) &= \psi_0(\mathbf{x}_\perp)\quad\text{and} \\
\psi_\text{L}(\mathbf{x}_\perp) &= \psi_0^*(\mathbf{x}_\perp),
\end{split}
\end{equation}
which implies that they have the same spatial profile but are traveling in opposite transverse directions, since complex conjugation reverses the sign of the transverse momentum phase factor.

What pattern of electrons will be observed in the observation plane? The probability density is computed by summing over the substrate modes:
\begin{equation}\label{eq:trace}
P_e(\mathbf{x}_\perp) = \sum_{n = 0,1} \braket{\Psi_f | \left( \mathbb{I}_\text{el} \otimes \ket{n}\bra{n} \right) | \Psi_f }.
\end{equation}
The wavefunction in Eq.~\ref{eq:simplified wavefunction} has four terms, so the sum in Eq.~\ref{eq:trace} yields sixteen terms, but we can simplify using orthogonality of modes (e.g., $\braket{0|1} = 0$, and $\braket{0|0}=1$), normalization (Eq.~\ref{eq:simplified normalization}), and the relationship between $\psi_\text{R}$ and $\psi_\text{L}$ (Eq.~\ref{eq:wavefunction overlap discussion}) to yield
\begin{equation}\label{eq:P_e(x)_partial_simplification}
\begin{split}
P_e(\mathbf{x}_\perp) = &|\psi_0(\mathbf{x}_\perp)|^2 \\
&+\frac{1}{2} (\psi_0(\mathbf{x}_\perp))^2 \left[ c^*_{\text{R}0} c_{\text{L}0} + c^*_{\text{R}1} c_{\text{L}1} \right] \\
&+\frac{1}{2} (\psi^*_0(\mathbf{x}_\perp))^2 \left[ c_{\text{R}0}c^*_{\text{L}0}  + c_{\text{R}1} c^*_{\text{L}1}  \right].
\end{split}
\end{equation}
This form reveals the structure of the calculation. The first line is an average ``background" term, and the others account for the possibility of an interference fringe.

\subsection{Fringe Visibility}\label{subsec:fringe_visibility}

Notice the structure of the probability density in Eq.~\ref{eq:P_e(x)_partial_simplification}. The second and third lines are complex conjugates of one another, so we can write this more neatly as 
\begin{equation}\label{eq:screen_probability_with_gamma}
P_e(\mathbf{x}_\perp) = |\psi_0(\mathbf{x}_\perp)|^2 +\mathrm{Re}\left[ \Gamma (\psi_0(\mathbf{x}_\perp))^2 \right]
\end{equation}
where the factor $\Gamma$ is the (possibly complex) value
\begin{equation}
\Gamma = c^*_{\text{R}0} c_{\text{L}0} + c^*_{\text{R}1} c_{\text{L}1}.
\end{equation}

\begin{figure}
\centering
\includegraphics[width=0.5\textwidth]{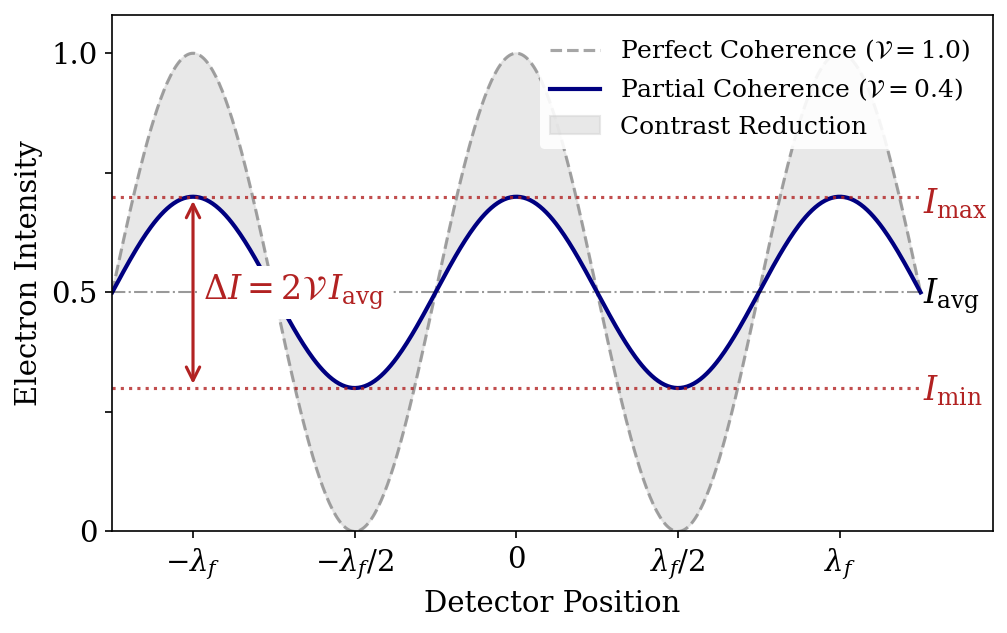}
\caption{Illustration of fringe visibility $\mathcal{V}$ in an interference pattern. The dashed curve shows the ideal case of perfect coherence $\mathcal{V}=1$, while the solid curve shows a partially coherent pattern with visibility $\mathcal{V}=0.4$. The average intensity $I_{\rm avg}$ is unchanged with the reduced visibility, but the maxima $I_{\rm max}$ and minima $I_{\rm min}$ are compressed toward the mean.}
\label{fig:visibility}
\end{figure}

The magnitude of $\Gamma$ is the fringe visibility $\mathcal{V}$: 
\begin{equation}
\mathcal{V} = |\Gamma|.
\end{equation}
Fig.~\ref{fig:visibility} illustrates schematically how the visibility reduction in an electron fringe appears experimentally. The peak-to-trough intensity difference is
\begin{equation}
\Delta I = I_{\rm max} - I_{\rm min} = 2\mathcal{V}I_{\rm avg},
\end{equation}
so decreasing the visibility $\mathcal{V}$ directly reduces the observable modulation of the fringes. 

Three limits for $\mathcal{V}$ are important to appreciate. Suppose that the mode coefficients for the right and left paths of the electron are equal, such that
\begin{equation}
c_{\text{R}0} = c_{\text{L}0} \quad \text{and} \quad c_{\text{R}1}=c_{\text{L}1}.
\end{equation}
In this case, we obtain $\mathcal{V} = 1$, with
\begin{equation}
P_e(\mathbf{x}_\perp) = |\psi_0(\mathbf{x}_\perp)|^2 + \text{Re}\left[ (\psi_0(\mathbf{x}_\perp))^2 \right], 
\end{equation}
which allows $P_e$ to vary from complete destructive interference to maximal constructive interference. This is the limit of \textit{full coherence}.

As a second possibility, suppose that the mode-path coefficients are unequal, but are related by 
\begin{equation}
c_{\text{R}0} = c_{\text{L}0}^{*}=c_0 \quad \text{and} \quad c_{\text{R}1}=c_{\text{L}1}^{*}=c_1.
\end{equation}
This gives us
\begin{equation}
P_e(\mathbf{x}_\perp) = |\psi_0(\mathbf{x}_\perp)|^2 + \text{Re}\left[ (c_0^2+c_1^2)(\psi_0(\mathbf{x}_\perp))^2 \right], 
\end{equation}
In general $c_0^2+ c_1^2$ is complex-valued, and its magnitude
\begin{equation}
\mathcal{V} = |c_0^2 + c_1^2| 
\end{equation}
lies in the range $0 < \mathcal{V} \leq 1$. This is one version of the most typical case, that of \textit{partial coherence}.

How would \textit{full incoherence} occur? Consider the (physically unlikely) case where the left path fully excites the substrate, while the right path completely fails to excite the substrate. Up to phase factors, this means
\begin{equation}
\begin{split}
c_{\text{R}0} = 1,\, c_{\text{R}1} &= 0, \\
c_{\text{L}0} = 0,\, c_{\text{L}1} &= 1,
\end{split}
\end{equation}
which is easy to confirm will lead to
\begin{equation}
P_e(\mathbf{x}_\perp) = |\psi_0(\mathbf{x}_\perp)|^2,
\end{equation}
i.e., to $\mathcal{V} = 0$, corresponding to the limit of \textit{full incoherence}, where no interference fringes are observable despite the spatial overlap of the wave packets.

\subsection{Summing over Multiple Modes}\label{subsec:multiple_modes}

Now let us revisit the case where the substrate has a countably infinite number of surface modes. The initial state of the system is still given by Eq.~\ref{eq:initial wavefunction}, but, working to first order in perturbation theory so that at most a single surface mode is excited, the final state after the electron has decoupled from the substrate may be written as
\begin{equation}
\begin{split}
\ket{\Psi_f}
=
\frac{1}{\sqrt{2}}
\Bigg[
&\psi_{\mathrm R}(\mathbf x_\perp)
\left(
c_{\mathrm R0}\ket{0}
+
\sum_{\mathbf k} c_{\mathrm R\mathbf k}\ket{1_{\mathbf k}}
\right)
\\
+
&\psi_{\mathrm L}(\mathbf x_\perp)
\left(
c_{\mathrm L0}\ket{0}
+
\sum_{\mathbf k} c_{\mathrm L\mathbf k}\ket{1_{\mathbf k}}
\right)
\Bigg],
\end{split}
\label{eq:Psi_f_heuristic}
\end{equation}
where $\ket{1_{\mathbf k}}$ denotes a single excitation of a surface mode labeled by the surface wavevector $\mathbf k = (k_x,0,k_z)$, and the coefficients satisfy the normalization conditions
\begin{equation}
|c_{j0}|^2+\sum_{\mathbf k}|c_{j\mathbf k}|^2=1,
\qquad j\in\{\mathrm R,\mathrm L\}.
\end{equation}

As above, the observed electron probability density is obtained by summing over the unobserved substrate degrees of freedom at the screen position as
\begin{equation}
P_e(\mathbf x_\perp) =
\sum_{\{n_{\mathbf k}\}}
\braket{
\Psi_f
|
\mathbb I_{\mathrm{el}}
\otimes
\ket{\{n_{\mathbf k}\}}\bra{\{n_{\mathbf k}\}}
|
\Psi_f
}.\label{eq:trace_heuristic}
\end{equation}
Here $\ket{\{n_{\mathbf k}\}}$ denotes a complete Fock basis of substrate excitation states, labeled by the occupation numbers of the surface modes $\mathbf k$. 

For the recombined beams, we presume the spatial profiles on the right and left for the overlap are related as
\begin{equation}
\psi_{\mathrm{R}}(\mathbf{x}_\perp) = \psi_0(\mathbf{x}_\perp) \quad \text{and} \quad \psi_{\mathrm{L}}(\mathbf{x}_\perp) = \psi_0^*(\mathbf{x}_\perp).
\end{equation}
At the same time, we relate the zero-loss transition coefficients on the right and left as
\begin{equation}
c_{\mathrm{R}0} = c_{\mathrm{L}0} = c_0.
\end{equation}
These choices correspond to a symmetric recombination of beams with opposite transverse momentum.

Using orthogonality relations (e.g., $\braket{0|1_{\mathbf k}}=0$ and
$\braket{1_{\mathbf k}|1_{\mathbf k'}}=\delta_{\mathbf k,\mathbf k'}$), 
the probability density reduces to
\begin{equation}
\begin{split}
P_e(\mathbf x_\perp) = &|\psi_0(\mathbf x_\perp)|^2
+ \\
&\mathrm{Re}
\Bigg\{
(\psi_0(\mathbf x_\perp))^2
\left[
c_{\mathrm R0}^*c_{\mathrm L0}
+
\sum_{\mathbf k}
c_{\mathrm R\mathbf k}^*c_{\mathrm L\mathbf k}
\right]
\Bigg\}.
\end{split}
\label{eq:P_pattern_heuristic}
\end{equation}
Comparing this with Eq~\ref{eq:screen_probability_with_gamma} above, we recognize that the coherence factor $\Gamma$ is
\begin{equation}
\Gamma
=
c_{\mathrm R0}^*c_{\mathrm L0}
+
\sum_{\mathbf k}
c_{\mathrm R\mathbf k}^*c_{\mathrm L\mathbf k},
\label{eq:Gamma_def}
\end{equation}
and, as above, the modulation depth of the interference pattern, will be determined by the magnitude of $\Gamma$:
\begin{equation}
\mathcal V = |\Gamma|,
\qquad 0 \le \mathcal V \le 1.
\end{equation}
In the present case $\Gamma$ is real, since the coordinates have been chosen in a symmetric way, so that $\mathcal V=\Gamma$.

\subsection{System Symmetry and Fringe Visibility}\label{subsec:symmetry}

To evaluate the \textit{right} and \textit{left} scattering coefficients, it is convenient to define a reference trajectory. Suppose $c_{\mathbf k}$ is the excitation coefficient that would arise for an unsplit beam traveling midway between the two paths along the central axis ($x=0$). By shifting the coordinate origin to the actual trajectories at $x = \pm d/2$, the coefficients for the right and left trajectories differ from the reference beam only by a geometric phase---specifically,
\begin{equation}
c_{\mathrm R \mathbf k} = e^{+i k_x d /2} c_{\mathbf k} 
\quad \text{and} \quad
c_{\mathrm L \mathbf k} = e^{-i k_x d /2} c_{\mathbf k}.
\end{equation}

Substituting into Eq.~\ref{eq:Gamma_def} gives
\begin{equation}\label{eq:Gamma_midway}
\Gamma
=
|c_{0}|^2
+
\sum_{\mathbf k}
|c_{\mathbf k}|^2 e^{-i k_x d}.
\end{equation}
The setup is symmetric under reflection across the central plane between the two trajectories (i.e., from $x \to -x$ across the $yz$ plate at $x = 0$), which means that the imaginary sine components cancel upon summation over all $\mathbf{k}$, leaving only the real cosine part. Using this and the first-order normalization condition
\begin{equation}
|c_{0}|^2+\sum_{\mathbf k}|c_{\mathbf k}|^2=1,
\end{equation}
we can therefore rewrite Eq.~\ref{eq:Gamma_midway} as
\begin{equation}
\mathcal{V}
=
1
-
\sum_{\mathbf k} |c_{\mathbf k}|^2 \left[ 1 - \cos(k_x d)\right].
\label{eq:V with cosine term}
\end{equation}

This quantifies how interference contrast is reduced by entanglement with substrate excitations, with coherence preserved by processes that leave the substrate in the same final state for both paths and suppressed by excitations that do not. For an extended interaction region, this overlap accumulates continuously, motivating the Markovian approximation developed below.

\subsection{Markovian Approximation}\label{subsec:Markov}

For an extended interaction region, we may use a Markovian approximation that imposes Poisson statistics for independent scattering events. The excitation probability grows linearly with propagation length as
\begin{equation}
|c_{\mathbf k}|^2
=
L\,\frac{\dd P_{\mathbf k}}{\dd z},
\end{equation}
where $\dd P_{\mathbf k}/\dd z$ is the excitation probability per unit length. Retaining terms linear in $L$ gives
\begin{equation}
\mathcal{V}(L)
\approx
1
-
L
\sum_{\mathbf k}
\frac{\dd P_{\mathbf k}}{\dd z}
\bigl[1-\cos(k_x d)\bigr].
\end{equation}
Assuming successive longitudinal slices are statistically independent, each slice will contribute to a small reduction in coherence, leading to a differential rate equation:
\begin{equation}
\frac{\dd \mathcal{V}}{\dd z}
=
-
\gamma(b,d)\,\mathcal{V},
\end{equation}
with
\begin{equation}
\gamma(b,d)
=
\sum_{\mathbf k}
\frac{\dd P_{\mathbf k}}{\dd z}
\bigl[1-\cos(k_x d)\bigr].
\label{eq:gamma_kernel}
\end{equation}

We can generalize this result by analogy. Suppose we have the differential EELS expression for $\dd^3 P/\dd \omega\,\dd k_x \,\dd z$. A generalized version of Eq.~\ref{eq:gamma_kernel} is
\begin{equation}\label{eq:gamma_continuum_restate}
\begin{aligned}
&\gamma(b,d)=\\
&\int_0^\infty \dd\omega
\int_{-\infty}^{\infty} \dd k_x \,\frac{\dd^3 P}{\dd \omega\,\dd k_x \, \dd z}\,\bigl[1-\cos(k_x d)\bigr],\end{aligned}
\end{equation}
which leads to an exponential loss in fringe visibility as
\begin{equation}\label{eq:fringe visibility}
\mathcal{V}(b,d,L) = \exp[-\gamma(b,d) L].
\end{equation}
This is the basic structure of each of the models discussed below. Eq.~\ref{eq:gamma_continuum_restate} makes explicit that decoherence is governed by the same differential loss spectrum that appears in EELS, weighted by how well a given mode distinguishes the two electron paths.

\subsection{Thermal Effects}\label{subsec:thermal effects}

Until now, all thermal effects have been ignored. But at finite temperature $T$, the surface modes are no longer in the vacuum state, but are thermally occupied according to Bose-Einstein statistics:
\begin{equation}
\bar{n}_{\mathbf k} =\frac{1}{e^{\hbar \omega_{\mathbf k}/k_B T}-1}.
\end{equation}

Each electron's interaction with these populated modes has two consequences: an emission channel in which the electron excites the mode and produces the usual factor $\bar{n}_{\mathbf k}+1$, and an absorption channel in which the electron removes a thermally populated quantum and produces a factor $\bar{n}_{\mathbf k}$. The total thermal weighting is thus
\begin{equation}
(\bar{n}_{\mathbf k}+1)+\bar{n}_{\mathbf k}=2 \bar{n}_{\mathbf k}+1
=\coth\!\left(\frac{ \hbar \omega_{\mathbf k}}{2 k_B T}\right).
\end{equation}
The finite-temperature decoherence rate is obtained from the $T=0$ expression by multiplying each mode contribution by this term \cite{Lucas1972, Idrobo2018}. 

An alternative way to approach the question of thermal occupancies is to introduce a substrate in state $\ket{n_\mathbf{k}}$ and to calculate the transition probabilities for that particular case, then to propose that the physically observed effect will result from a thermally weighted (classical) average of these interactions. Such an analysis recovers this same factor of $\coth(\hbar \omega_\mathbf{k}/2 k_B T)$.

This thermal weighting implies in general that the decoherence weight for each mode needs to be weighted as
\begin{equation}
\begin{aligned}
&\gamma(b,d,T)= \\
&\int_0^\infty \dd\omega
\int_{-\infty}^{\infty} \dd k_x \,\frac{\dd^3 P}{\dd \omega\,\dd k_x \, \dd z}\,\bigl[1-\cos(k_x d)\bigr] \coth\!\left(\frac{\hbar\omega}{2k_B T}\right),
\end{aligned}
\label{eq:gamma_continuum_restate_thermal}
\end{equation}
which, as before, can generate experimental predictions for the fringe visibility via $\mathcal{V} = \exp(-\gamma(b,d,T) L)$.

\begin{figure}
\centering
\includegraphics[width=.5\textwidth]{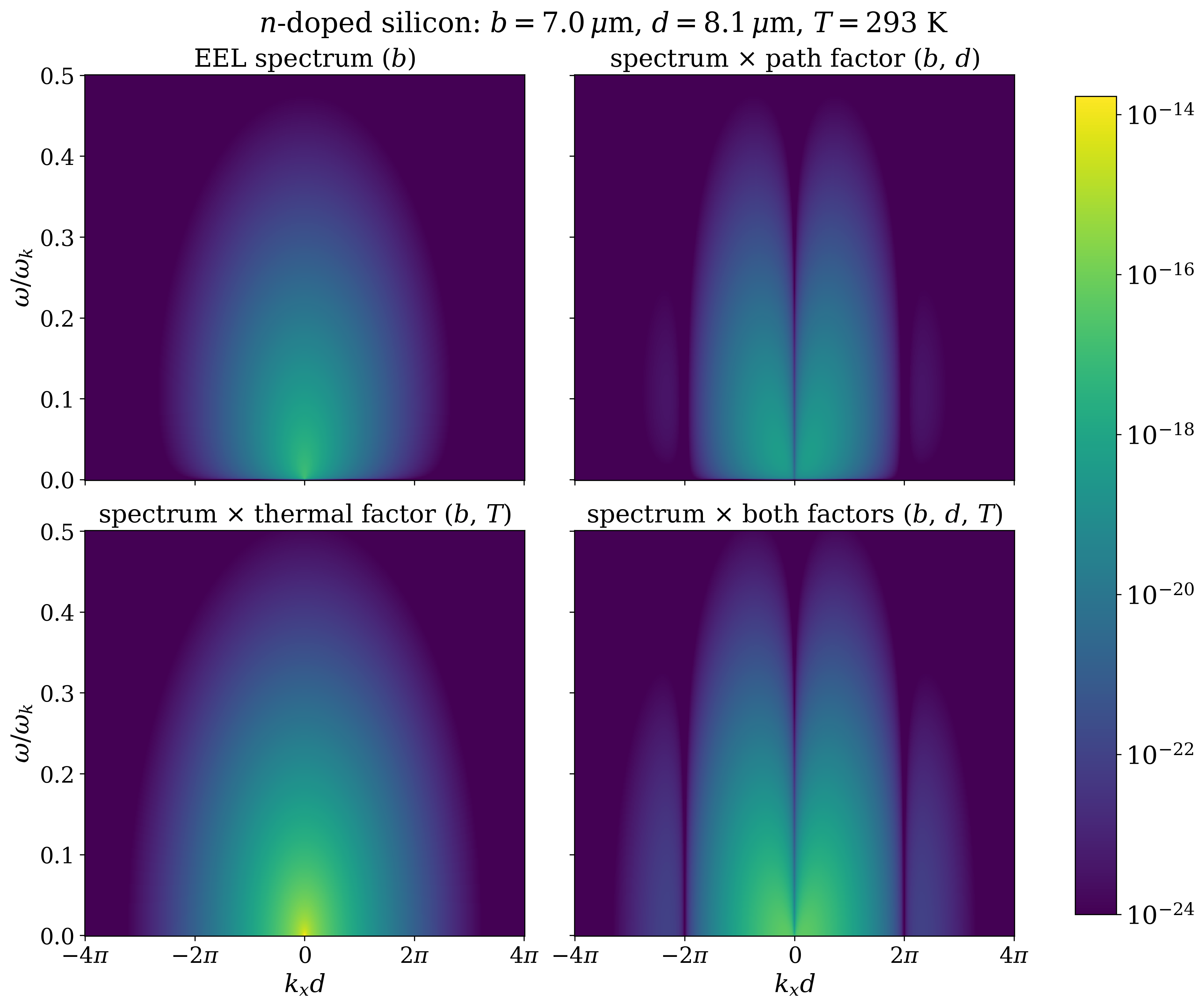}
\caption{Mode-resolved contributions to the decoherence rate integrand in the $(k_x,\omega)$ plane for an electron beam at height $b = 7.0\,\mu$m above an \textit{n}-doped silicon surface, with path separation $d=8.1\,\mu$m, and at temperature $T = 293$ K. (Model specifics are given in Sec.~\ref{sec:calculation} below.) Top left: differential EEL spectrum $\dd^3 P/(\dd z\,\dd \omega\,\dd k_x)$. Top right: product of the EEL spectrum with the path weighting factor $1-\cos(k_x d)$. Bottom left: product of EEL spectrum with the thermal factor $\coth(\hbar\omega/2k_B T)$. Bottom right: the full integrand entering $\gamma(b,d,T)$---the differential EEL spectrum multiplied both by the path and the thermal weighting.}\label{fig:decoherence-integrand}
\end{figure}

Fig.~\ref{fig:decoherence-integrand} illustrates how the integrand in Eq.~\ref{eq:gamma_continuum_restate_thermal} combines geometric, interference, and thermal effects. The impact parameter $b$ enters through the differential EEL spectrum $\dd^3 P/ \dd \omega \, \dd k_x \, \dd z$ via an evanescent factor that suppresses coupling to modes with large lateral wavevector. The path separation $d$ appears in the factor $1-\cos(k_x d)$, which produces oscillatory zeros at $k_x d = 2\pi n$. Finally, the temperature $T$ enters through the factor $\coth(\hbar\omega/2k_B T)$, which enhances low-frequency modes and increases their contribution to the decoherence rate. 

From this, we may already draw some broad conclusions. The dominant contributions to $\gamma(b,d,T)$ arise from those modes that are sufficiently long-wavelength to reach the beam at height $b$, with lateral wavevector large enough to distinguish the two paths, and with frequencies low enough to be thermally populated.

\section{Visibility Loss Models}\label{sec:calculation}

For numerical predictions, the theoretical framework above must be informed both with experimental parameters (e.g., with the $v$, $b$, $d$, and $L$ values as labeled in Fig.~\ref{fig:substrate-geometry}), as well as material models for the dielectric substrate that feed into the differential EELS models. One advantage of this option is that we can choose to inform our EELS models with different physics to examine which effects are observationally most significant. Of course, constructing such EELS models takes an extra step. Here we quote the relevant results. Such results are themselves semi-classical, but for those readers who are interested in a quantized approach we have derived the simplest case (see Sec.~\ref{subsec:lossless_model} below) as Appendix~\ref{sec:scattering}.

In Sec.~\ref{sec:experiment}, we compare calculations with experimental results from Kerker \textit{et al.}  \cite{Kerker-etal2020}, so here we will quote the model that they used to capture material excitations. They studied the decohering properties of both gold and $n$-doped silicon, each of which can (at low frequencies) be modeled by a three-parameter Drude model with background dielectric constant:
\begin{equation}
\varepsilon(\omega) = \epsilon_L \left( 1  - \frac{\omega_{\rm p}^2}{\omega(\omega + \ii\gamma_{\rm rel})} \right).
\label{eq:drude_model}
\end{equation}
This form reflects that the dielectric response is dominated at low frequencies by the static lattice contribution $\epsilon_L$, while the motion of free charges is captured by $\omega_{\rm p}$. The parameter $\gamma_{\rm rel}$ quantifies the rate of mode relaxation. 

Using this concrete model of $\varepsilon(\omega)$, it is straightforward to verify that each of the models of decoherence below reduces to the one discussed immediately before it.

\subsection{Lossless model without signal retardation}\label{subsec:lossless_model}

First we consider the limiting case of electrostatic coupling ($\omega/c \rightarrow 0$) with undamped surface modes, for which the Drude-Lorentz model is used for $\varepsilon(\omega)$ but with $\gamma_{\rm rel} = 0$. The modes are each labeled by their wavevector $\mathbf{k}$. From the electrostatic boundary condition, their oscillation frequency $\omega_\mathbf{k}$ must satisfy
\begin{equation}
\varepsilon(\omega_\mathbf{k}) = -1,
\label{eq:SPP condition}
\end{equation}
the familiar condition for nonretarded surface plasmons at a planar interface with the vacuum.\cite{Pitarke2007} This can be inverted to find $\omega_\mathbf{k}$: 
\begin{equation}
\omega_{\mathbf{k}} = \omega_\mathrm{p} \sqrt{\frac{\epsilon_L}{\epsilon_L + 1}}.
\label{eq:lossless_omega}
\end{equation}

We have included a full derivation of this case in Appendix~\ref{sec:scattering}, including explicitly quantized modes.

This model severely underestimates the observed loss of visibility $\mathcal{V}$ in all regions, but it has the benefit of being analytically tractable. In the end, $\gamma(b,d)$ can be expressed in terms of modified Bessel functions as
\begin{equation}\label{eq:gamma_lossless}	
\begin{split}
&\gamma(b,d)
=\\
&\frac{2\alpha c \omega_{\mathbf k}}{(\epsilon_L+1)\,v^2}\,
\left[
K_0\!\left(\frac{2b\omega_{\mathbf k}}{v}\right)
-
K_0\!\left(
\frac{\omega_{\mathbf k}}{v}\sqrt{(2b)^2+d^2}
\right)
\right],
\end{split}
\end{equation}
where $\alpha \approx 1/137$ is the fine structure constant, $v$ is the electron speed, and $c$ is the speed of light. 

A larger $\gamma$ leads to a quicker reduction in the fringe visibility $\mathcal{V}$, so despite its limitations this model captures the relevant qualitative effects. It is notable that as the path separation $d \rightarrow 0$, the visibility $\mathcal{V} \rightarrow 1$. Furthermore, when $d \rightarrow \infty$, $\gamma(b,d)$ simply reduces to $\dd P/\dd z$. 

\subsection{Damped model without signal retardation}\label{subsec:damped model without signal retardation}

A more general model allows us to incorporate mode damping into the physical description (i.e., $\gamma_{\rm rel} \neq 0$), while still ignoring magnetic effects from signal retardation (i.e., $\omega/c \rightarrow 0$). When an electron travels in vacuum at a distance $b$ above a planar dielectric half-space, the differential energy loss expression in the nonretarded limit \cite{EcheniquePendry_1975} is
\begin{equation}
\frac{\dd^3P}{\dd\omega \,\dd k_x \,\dd z}
=
\frac{\alpha \, c}{\pi\,v^2}\;
\frac{\ee^{-2k_\parallel b}}{k_\parallel}\;
\Im\!\left[
\frac{\varepsilon(\omega)-1}{\varepsilon(\omega)+1}
\right],
\label{eq:dielectric_spectrum}
\end{equation}
where
\begin{equation}
k_\parallel =\sqrt{k_x^2 + \left(\omega/v\right)^2},
\label{eq:k_parallel}
\end{equation}
and $\varepsilon(\omega)$ is the complex dielectric function of the substrate. Plugging this into Eq.~\ref{eq:gamma_continuum_restate} and performing the integral over $k_x$ yields a semi-analytic expression for $\gamma(b,d)$:
\begin{equation}\label{eq:gamma_nonretarded}
\begin{split}
\gamma(b,d)
=\qquad&\\
\frac{2 \alpha c}{\pi\,v^2}
\int_0^\infty \dd\omega &\;
\Im\!\left[
\frac{\varepsilon(\omega)-1}{\varepsilon(\omega)+1}
\right] \times \\
&\left[
K_0\!\left(\frac{2b\omega}{v}\right)
-
K_0\!\left(
\frac{\omega}{v}\sqrt{(2b)^2+d^2}
\right)\right].
\end{split}
\end{equation}
If we expand $\varepsilon(\omega)$ around $-1$, Eq.~\ref{eq:gamma_nonretarded} reduces to the lossless model for $\gamma(b,d)$, Eq.~\ref{eq:gamma_lossless}, in the limit that $\gamma_{\rm rel} \rightarrow 0$. 

\subsection{Damped model with signal retardation}

The same logic can be used to generate predictions from a model that includes retardation. Retardation is incorporated through the functions
\begin{equation}\label{eq:nu_and_nu0}
\begin{split}
\nu(\omega,\,k_x) &= \sqrt{k_\parallel^2 - \varepsilon(\omega) \frac{\omega^2}{c^2}}\qquad\text{and} \\
\nu_0(\omega,\,k_x) &= \sqrt{k_\parallel^2 - \frac{\omega^2}{c^2}},
\end{split}
\end{equation}
where $\varepsilon(\omega)$ is the complex dielectric function of the substrate and $k_\parallel = \sqrt{k_x^2+(\omega/v)^2}$ as before. In the retarded formulation, the differential EELS expression \cite{Garcia-Molina_1985} is
\begin{equation}
\frac{\dd^3P}{\dd\omega\,\dd k_x\,\dd z}
=
\frac{\alpha \, c}{\pi v^2}\,
\frac{\ee^{-2\nu_0 b}}{\nu_0}\,
\Im\!\left[
\lambda_{\rm el}(\omega,\, k_x)
\right],
\label{eq:retarded_spectrum}
\end{equation}
where the loss function $
\lambda_{\rm el}(\omega,\,k_x)$ is given by
\begin{equation}\label{eq:lambda_retarded}
\begin{split}
&\lambda_{\rm el}(\omega, k_x) = \qquad \\
& \frac{1}{\nu_0 + \nu} \left( \frac{2 \nu_0^2 (\varepsilon - 1)}{\varepsilon \nu_0 + \nu} - \left(1 - \frac{v^2}{c^2} \right)(\nu_0 - \nu) \right).
\end{split}
\end{equation}
The retarded loss expression $\Im[\lambda_{\rm el}]$ reduces to the non-retarded expression
$\Im[(\varepsilon-1)/(\varepsilon+1)]$ in the limit $\omega/c\rightarrow 0$.

This retarded, thermally weighted EELS model is the central result of the present derivation. Substituting Eq.~\ref{eq:retarded_spectrum} into the thermally weighted decoherence rate $\gamma(b,d,T)$ (Eq.~\ref{eq:gamma_continuum_restate_thermal}) recovers the macroscopic QED model of Scheel and Buhmann in its Markovian limit\cite{ScheelBuhmann2012}, the model Kerker \textit{et al.} identified as giving the best agreement with their measured fringe visibilities.\cite{Kerker-etal2020} To our knowledge this connection has not been previously reported. It shows that the entanglement-based decoherence framework developed above, following Strauch's pedagogical treatment,\cite{Strauch2025} is not just qualitatively suggestive but reduces exactly to the specific model already validated against experiment.

\section{Experimental Implications}\label{sec:experiment}

Now we can compare models directly with experimental results from Kerker \textit{et al.} The Kerker \textit{et al.} experiments employed a coherent 1~keV electron beam (with speed $v \approx 0.062 c$) propagating parallel to a surface of length $L = 0.01$ m over either \textit{n}-doped silicon or gold, with the fringe visibility measured as a function of beam height $b$ for several fixed lateral path separations $d$. 

\begin{figure}
\centering
\includegraphics[width=0.5\textwidth]{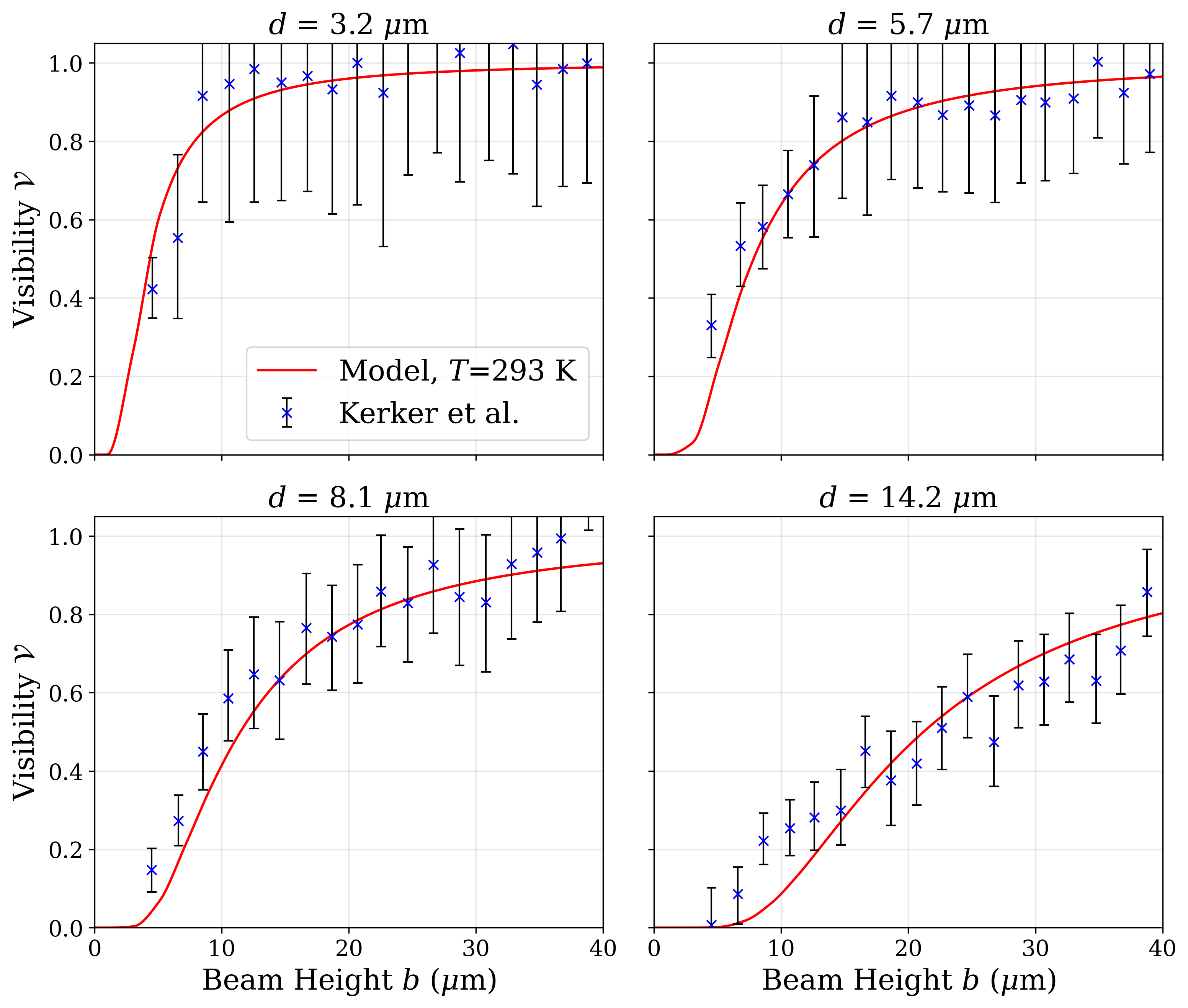}
\caption{Visibility $\mathcal{V}$ as a function of surface distance $b$ for several fixed path separations $d$ (indicated in each panel), for a 1 keV electron passing over \textit{n}-doped silicon. Discrete points show digitized results from plots in Kerker \textit{et al.}, and solid curves correspond to results from the retarded model (Eq.~\ref{eq:retarded_spectrum}) applied to the decoherence rate with a $T=293$ K thermal correction (Eq.~\ref{eq:gamma_continuum_restate_thermal}).}\label{fig:experimental-comparison_silicon}
\end{figure}

Here we will restrict our attention to the \textit{n}-doped silicon. For the modified Drude model (Eq.~\ref{eq:drude_model}) of an \textit{n}-doped silicon surface, we use the parameters
\begin{equation}
\begin{split}
\epsilon_L &= 11.7,\\
\omega_{\rm p} &= 2.48 \times 10^{13} \,\text{rad/s, and} \\
\gamma_{\rm rel} &=2.59 \times 10^{12} \,\text{rad/s}.
\end{split}\label{eq:drude_parameters}
\end{equation}
These parameters are taken from Karstens \cite{Karstens2014}, following the analysis of Kerker \textit{et al.}~\cite{Kerker-etal2020}. Fig.~\ref{fig:experimental-comparison_silicon} compares the resulting predictions with visibility data $\mathcal{V}$ digitized from Kerker \textit{et al.}, using their reported experimental geometry and distance calibration. As in that publication, an additional $3~\mu\mathrm{m}$ offset has been added to the values of $b$ to correct for a reported systematic error in the data. 

The plots in Fig.~\ref{fig:experimental-comparison_silicon} show $\mathcal{V}$ as a function of $b$ for a split-path electron passing over \textit{n}-doped silicon, with each subplot corresponding to a fixed path separation $d$. Visibility increases quickly with increasing beam distance from the interface $b$, and decreases with increasing path separation $d$. This is just as we might expect. Entanglement between the probe and the substrate depends on how near they are to one another during their interaction. Likewise, the farther separated the electron paths, the more distinguishable their interactions with the substrate become. It is worth emphasizing that the theoretical curves here are not empirical fits, but the direct output of the macroscopic QED model of Scheel and Buhmann, the endpoint of the derivation developed above. The agreement therefore illustrates how a first-principles model, built from basic quantum considerations, can confront real experimental data without adjustable parameters.

\begin{figure}
\centering
\includegraphics[width=0.5\textwidth]{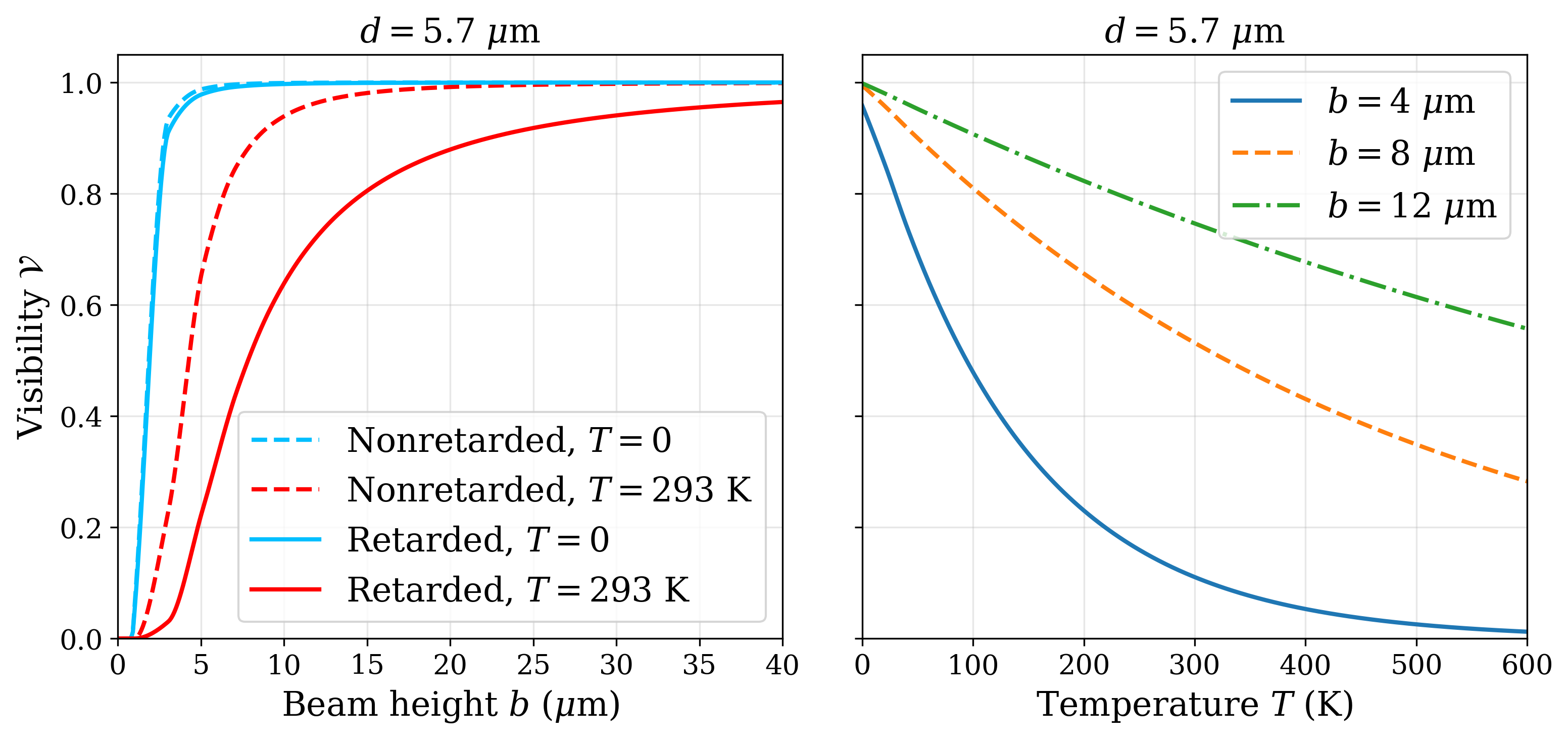}
\caption{Fringe visibility $\mathcal{V}$ for 1 keV electrons passing over $n$-doped silicon ($d = 5.7\,\mu\mathrm{m}$). \textit{Left}: Model comparison showing that both retardation (solid curves) and thermal occupation at $T=293$ K (red curves) are necessary to reproduce the experimental data in Fig.~\ref{fig:experimental-comparison_silicon}. Dashed curves show non-retarded limits, while blue curves show $T \to 0$ limits. \textit{Right}: Temperature dependence of $\mathcal{V}$ at three impact parameters $b$, computed using the fully retarded, thermally weighted model.}\label{fig:V_vs_T}
\end{figure}

Fig.~\ref{fig:V_vs_T} demonstrates that such models can do more than simply reproduce experimental results. The left subplot of Fig.~\ref{fig:V_vs_T} shows how calculations may isolate the relative importance of various effects. For instance, despite the relatively low-energy electron beam, non-retarded EELS models (dashed curves) significantly underestimate the visibility loss relative to retarded models (solid curves). Furthermore, non-thermal calculations (blue curves, $T \rightarrow 0$) also underestimate the loss of $\mathcal{V}$ relative to thermal calculations (red curves, calculated with the $\coth\!\left(\hbar\omega/2k_B T \right)$ with $T = 293$ K). Both retardation and thermal occupation effects are necessary to reproduce the observed scale of the visibility loss. 

The right subplot of Fig.~\ref{fig:V_vs_T} suggests an experimental extension based on this framework. As Velasco \textit{et al.} have explored,\cite{Velasco2024, Velasco2026} fringe visibility depends sensitively on substrate temperature, since the thermal factor $\coth(\hbar\omega/2k_BT)$ enters the decoherence rate as a multiplicative weight on the loss spectrum. From low temperatures to 600 K, visibility varies strongly with both $T$ and $b$. This raises the possibility of using fringe visibility as a non-invasive thermal probe for nanoscale systems.

Kerker \textit{et al.} also included results for a gold surface, for which $\mathcal{V}$ increases much more quickly with increased distance $b$ than for the \textit{n}-doped silicon, but indicated further experiments are needed to validate results at small $b$. Intriguingly, the retarded ``correction" terms can be dominant for determining decoherence rates in metals.\cite{Howie2019} Unlike in silicon, the $\gamma(b,d,T)$ expression for gold has an apparent infrared divergence, requiring a low-frequency cutoff in the integral based on the time of flight of the electron over the surface.\cite{Velasco2026} We refer readers to the published sources for further discussions.

\section{Conclusion}

We have shown how a schematic model can be augmented by a small number of physically motivated steps (incorporating multiple surface modes, applying a Markov approximation, including thermal occupation effects) to recover an empirically predictive theory. Framing this connection through a mode-resolved scattering picture ties the decoherence rate directly to familiar models from electron energy-loss spectroscopy. This mode-resolved picture also clarifies why temperature matters, since the surface-mode energies are low enough that significant thermal population occurs even at room temperature. It further implies that fringe visibility measurably varies with temperature, suggesting applications as a thermal probe. 

More broadly, this discussion illustrates a useful pattern for physics pedagogy. Idealized toy models can often be extended by a small number of well-motivated steps into tools with genuine predictive power.

\begin{acknowledgments}
The author thanks Archie Howie for clarifying comments and Robin R\"{o}pke for help with numerics.
\end{acknowledgments}

\appendix

\section{From surface modes to EELS}\label{sec:scattering}

This appendix outlines the calculation of quantum scattering amplitudes coupling an electron beam to surface modes of a dielectric half-space, working its way up to a description of an EEL spectrum.

The quantization of surface plasmon polariton modes and their excitation by external charges is well-established in both semiclassical and quantum frameworks \cite{Lucas-etal1970, Wang1996, Zhang_2012}. It has been applied to many regular geometries in the nonretarded ($\omega/c \to 0$) limit \cite{deAbajo2010,Kordahl2019}. Substrate dynamics are encoded through the relative permittivity $\varepsilon(\omega)$, which is generally complex, but for the lossless mode construction, we take $\varepsilon(\omega)$ to be real, so that the field can be decomposed into normal modes. 

\subsection{Electrostatic Surface Modes}

For the dielectric half-space, we introduce mode potentials
\begin{equation}\label{eq:modes}
\phi_\mathbf{k}(\mathbf{r}) = \frac{1}{\sqrt{L^2 k}} \begin{cases}
e^{i \mathbf{k} \cdot \mathbf{r}} e^{+ky},\quad \text{for} \quad y < 0; \\
e^{i \mathbf{k} \cdot \mathbf{r}} e^{-ky},\quad \text{for} \quad y \geq 0,
\end{cases}
\end{equation}
whose wave vector $\mathbf{k}$ lies parallel to the $xz$ plane as
\begin{equation}
\begin{split}
\mathbf{k} &= (k_x, 0, k_z) \text{ and } k = \sqrt{k_x^2 + k_z^2}.
\end{split}
\end{equation}
Though this expression includes $L$ as the box normalization length, all physical observables will be independent of this choice.

Defining the associated electric field modes as
\begin{equation}
\mathbf{u}_\mathbf{k}(\mathbf{r}) = \mathbf{\nabla} \phi_\mathbf{k}(\mathbf{r})
\end{equation}
the modes are normalized such that integrating over the dielectric half-space yields
\begin{equation}
\int_\mathrm{y<0} \mathrm{d}^3 \mathbf{r} \, \mathbf{u}_\mathbf{k}(\mathbf{r}) \cdot \mathbf{u}^*_\mathbf{k'}(\mathbf{r}) = \delta_{\mathbf{k},\mathbf{k'}}.
\end{equation}

At the planar interface, continuity of the normal component of the electric displacement requires
\begin{equation}
\varepsilon_r(\omega)\,
\hat{\mathbf n}\!\cdot\!\mathbf u_{\mathbf k}(\mathbf r)\big|_{y\to 0^-}
=
\hat{\mathbf n}\!\cdot\!\mathbf u_{\mathbf k}(\mathbf r)\big|_{y\to 0^+}.
\end{equation}
For the dielectric half-space, this implies that
\begin{equation}
\varepsilon_r(\omega_\mathbf{k}) = -1,
\end{equation}
within the dielectric for all $\mathbf{k}$, which can be inverted to find $\omega_\mathbf{k}$ (cf. Sec.~\ref{subsec:lossless_model} above). 

\subsection{Quantization and Normalization}

For a classical potential mode $\Phi_\mathbf{k}(\mathbf{r})$ with physical units, the time-averaged electric field energy (cf. Landau and Lifshitz\cite{landau_lifshitz_edcm}, \S 80) is
\begin{equation}\label{eq:classical_mode_energy}
U_\mathbf{k} = \frac{\epsilon_0}{2} \int \mathrm{d}^3\mathbf{r} \,\partial_\omega [\omega \varepsilon_r(\mathbf r,\omega)]_{\omega = \omega_{\mathbf{k}}}
   |\nabla \Phi_\mathbf{k}(\mathbf r)|^2,
\end{equation}
where the integral is over all space, and $\varepsilon_r(\omega) = 1$ in vacuum. For our normalized potentials, we may calculate
\begin{equation}
N_\mathbf{k} = \frac{1}{2} \int \mathrm{d}^3\mathbf{r} \,\partial_\omega [\omega \varepsilon_r(\mathbf r,\omega)]_{\omega = \omega_{\mathbf{k}}}
   |\nabla \phi_\mathbf{k}(\mathbf r)|^2,
\end{equation}
which, given the above conventions, reduces to 
\begin{equation}\label{eq:normalization nk}
N_\mathbf{k} = \frac{1}{2} \left( \partial_\omega [\omega \varepsilon_r(\omega)]_{\omega = \omega_{\mathbf{k}}} + 1 \right),
\end{equation}
with $\varepsilon_r(\omega)$ now as the relative permittivity of the substrate, and the ``+1'' term arising from the integral over the vacuum half-space. We use $N_\mathbf{k}$ here to avoid confusion with the thermal occupation numbers $n_\mathbf{k}$.

We now introduce creation and annihilation operators $\hat{a}_{\mathbf{k}}^\dagger$ and $\hat{a}_{\mathbf{k}}$ satisfying $[\hat{a}_{\mathbf{k}}, \hat{a}_{\mathbf{k'}}^\dagger] = \delta_{\mathbf{k}\mathbf{k'}}$. The free Hamiltonian for these modes takes the form of a quantum harmonic oscillator:
\begin{equation}
\hat{H}_\mathrm{free} = \sum_\mathbf{k} \hbar \omega_\mathbf{k} \left (\hat{a}_\mathbf{k}^\dagger\hat{a}_\mathbf{k} + \frac{1}{2} \right).
\end{equation}
To connect our classical fields to this quantum Hamiltonian, we must properly normalize the potential operator. The quantized potential operator bearing physical units is
\begin{equation}
\hat{\Phi}_\mathbf{k}(\mathbf{r}) = \sqrt{\frac{\hbar \omega_\mathbf{k}}{2\epsilon_0 N_\mathbf{k}}} \left( \phi^*_\mathbf{k} (\mathbf{r}) \hat{a}^\dagger_\mathbf{k} + \phi_\mathbf{k} (\mathbf{r}) \hat{a}_\mathbf{k} \right).
\end{equation}
The interaction Hamiltonian of these modes with a point charge $q$ following the prescribed trajectory $\mathbf{r}_q(t)$ is then
\begin{equation}
\hat{H}_\mathrm{int}(t) = \sum_\mathbf{k} q \hat{\Phi}_\mathbf{k}(\mathbf{r}_q(t)).
\end{equation}

\subsection{Aloof Energy-Loss Spectrum}

Using this interaction Hamiltonian, we now calculate the electron energy-loss spectrum for an electron traveling parallel to the surface, in the ``aloof'' geometry. The electron moves along the trajectory $\mathbf{r}_q(t) = (0, b, vt)$.

The first-order $\ket{0} \rightarrow \ket{1_\mathbf{k}}$ transition amplitude (denoted as $c_{\mathbf k}$ in the main text) is 
\begin{equation}
c_{0 \rightarrow 1_{\mathbf{k}}} = -\frac{i}{\hbar} \int_{t_{-}}^{t_{+}} \mathrm{d}t \bra{1_\mathbf{k}} \hat{H}_\mathrm{int}(t) e^{i \omega_{\mathbf{k}} t} \ket{0},
\end{equation}
where the integral limits span the time of one box length (i.e., $t_{-} = -L/2v$ to $t_{+} = +L/2v$, the time taken to traverse one quantization length). This leads to
\begin{equation}\label{eq:0->1 transition coefficient, no split}
c_{0 \rightarrow 1_{\mathbf{k}}} = -\frac{i q}{\hbar v} \sqrt{\frac{\hbar \omega_\mathbf{k}}{2\epsilon_0 N_\mathbf{k}}} \frac{e^{- k b}}{\sqrt{k}} \text{sinc}\left( \frac{(\omega_\mathbf{k} - k_z v) L}{ 2 v} \right), 
\end{equation}
where $\text{sinc}(x) \equiv \sin(x) / x$. In the large $L$ limit, the squared modulus of $c_{0 \rightarrow 1_{\mathbf{k}}}$ becomes a delta function:
\begin{equation}
\begin{split}
|c_{0 \rightarrow 1_{\mathbf{k}}}|^2 &= \frac{\pi q^2}{\epsilon_0 \hbar v^2} \frac{ \omega_\mathbf{k}}{ N_\mathbf{k}} \frac{e^{-2 k b}}{L k} \delta(k_z - \omega_\mathbf{k}/v). 
\end{split}\label{eq:pre-EELS probability}
\end{equation}
This is the probability of scattering for a single mode.

To find the probability of scattering from any $\phi_\mathbf{k}$ mode, we evaluate sums as integrals and add up the probabilities:
\begin{equation}
\begin{split}
\sum_{\mathbf{k}} P_{0 \rightarrow 1_\mathbf{k}} &= \frac{L^2}{(2\pi)^2} \int dk_x \, dk_z |c_{0 \rightarrow 1_{\mathbf{k}}}|^2 \\
&= \frac{q^2 L}{2\pi \epsilon_0 \hbar  v^2} \frac{\omega_\mathbf{k}}{N_\mathbf{k}} K_0 \left( \frac{2 \, b \, \omega_\mathbf{k}}{v} \right),
\end{split}
\end{equation}
where $K_0(x)$ is the modified Bessel function of the second kind. If the passing charge is an electron ($q = -e$), this result can be written in terms of the fine structure constant $\alpha \approx 1/137$ and the speed of light $c$ as
\begin{equation}\label{eq:surface scattering probability - no split}
\sum_{\mathbf{k}} P_{0 \rightarrow 1_\mathbf{k}} = 2 \alpha c \frac{L}{v^2} \frac{\omega_\mathbf{k}}{N_\mathbf{k} } K_0 \left( \frac{2 \, b \, \omega_\mathbf{k}}{v} \right).
\end{equation}
The $K_0(2b\omega/v)$ term here expresses the exponential suppression of the excitation probability with increasing distance between the beam and the substrate.

The linear dependence of the total probability on $L$ suggests an interpretation in terms of a probability per unit length. If we wish, we can rewrite Eq.~\ref{eq:pre-EELS probability} as an EEL spectrum, to be integrated over $\omega$ and $k_x$, as 
\begin{equation}
\frac{\dd^3P}{\dd\omega\,\dd k_x\,\dd z}
= \alpha c \frac{\omega}{v^2 N_\mathbf{k}}\,
\frac{e^{-2 b k_\parallel}}{k_\parallel}\,
\delta(\omega-\omega_\mathbf{k})
\label{eq:undamped EELS}
\end{equation}
where, as above, $k_\parallel \equiv \sqrt{k_x^2+(\omega/v)^2}$. This result coincides with the semiclassical EELS expression in the limit of vanishing damping \cite{Howie1985}.

\bibliography{bibliography}

\end{document}